\documentclass[preprint,superscriptaddress]{revtex4-1}
\usepackage{graphicx}
\usepackage{amsmath}
\usepackage{amsfonts}
\usepackage{amssymb}
\usepackage[utf8]{inputenc}

\begin{document}

\title{Role of Coulomb interaction in elastic pion-proton scattering from holography}

\author{Yu-Peng Zhang}
\email{zhangyupeng0108@foxmail.com}
\affiliation{School of Nuclear Science and Technology, University of South China, \\Hengyang 421001, People's Republic of China}

\author{Xun Chen}
\email{chenxunhep@qq.com}
\affiliation{School of Nuclear Science and Technology, University of South China, \\Hengyang 421001, People's Republic of China}

\author{Xiao-Hua Li}
\email{lixiaohuaphysics@126.com}
\affiliation{School of Nuclear Science and Technology, University of South China, \\Hengyang 421001, People's Republic of China}

\author{Akira Watanabe}
\email{watanabe@usc.edu.cn (Corresponding author)}
\affiliation{School of Mathematics and Physics, University of South China, \\Hengyang 421001, People's Republic of China}

\date{\today}

\begin{abstract}
Differential cross sections of the elastic pion-proton scattering are investigated at very small momentum transfer in a holographic QCD model, considering both the strong and Coulomb interaction in the Regge regime. 
The strong interaction is described by the Pomeron and Reggeon exchange, and the Coulomb interaction is characterized by the one photon exchange.
The two interactions are linked through an interference term and we only need to determine a single adjustable parameter involved in this term.
As to the parameters for the strong interaction, we can utilize the values determined in the previous studies.
The differential cross sections can be predicted without any additional parameters, and it is shown that our predictions are consistent with the experimental data.
We explicitly show the momentum transfer dependence for the interference effect.
The energy dependence of the contribution ratios for each component is also discussed.
\end{abstract}

\maketitle

\section{Introduction}
\label{sec:introduction}
The two-body elastic scattering of hadrons is one of the simplest processes, and has played an important role in investigating the partonic structure of the involved hadrons for several decades.
However, it still presents a challenge for the theory to describe it. 
The differential cross section is a fundamental experimental characterization determined by studying the elastic scattering involving hadrons.
In the present work we investigate the differential cross section of elastic pion-proton ($\pi p$) scattering with relatively high center-of-mass energy $s$ and quite low momentum transfer $t$.
Among the fundamental forces associated with hadron interactions, two fundamental interactions have commonly been employed to describe the differential cross section of two charged hadrons: the short-ranged strong interaction and the relatively weak but long-ranged Coulomb interaction.

Quantum chromodynamics (QCD) is a well-established theory of the strong interaction, and all phenomena associated with strong interactions are expected to be described by the fundamental QCD Lagrangian.
Since before the establishment of QCD, there has been a time-honored theory, the Regge theory, which is a theory for the analytic properties in high energy scattering as a function of angular momentum, where the angular momentum is not restricted to integer multiples, but is allowed to take any complex values. 
The Regge theory is a useful framework for analyzing the properties of scattering amplitudes for various high energy forward scattering processes~\cite{in1,in2,in3,in4,in5,in6,in7}.
The Regge theory successfully describes the proton-proton $(pp)$ and proton-antiproton $(p\Bar{p})$ scattering by considering the soft Pomeron and Reggeon exchange~\cite{in8,in9,in10,in11}, with their respective slope and intercept parameters.
Subsequently, the birth of the string theory was originally intended to unravel the modes of the strong interaction, and it is intimately linked with the Regge theory, where the S-matrix in the Regge theory can be described by the bosonic strings and the string amplitude can explicitly reproduce the Regge behavior for various cross sections.
Later on, QCD replaced the string theory as the fundamental theory of the strong interaction.
However, low momentum transfer characterizes hadronic elastic scattering, and the coupling of QCD becomes large at low energy scales. 
Therefore these scattering cross sections in QCD are essentially nonperturbative physical quantities that are difficult to analyze directly using QCD. 
In some quite limited kinematic region, perturbative QCD is available and important results have been obtained~\cite{in12,in13,in14}, but for most of the kinematic region we need effective methods to calculate the cross sections.
Over the past decades, instead of describing the elastic scattering from first principles, various theoretical models~\cite{in15,in16,in17,in18,in19,in20,in21,in22} have been developed to elucidate the experimental results for the elastic scattering.

In this work we consider the strong interaction in the framework of holographic QCD, a nonperturbative method for QCD that has been established using the anti-de Sitter/conformal field theory (AdS/CFT) correspondence~\cite{in23,in24,in25,in26}, which correlates a four-dimensional conformal field theory with the theory of gravity in the higher dimensional AdS space. 
Based on the AdS/CFT correspondence, many holographic models~\cite{Gursoy:2007cb,Gursoy:2007er,Gursoy:2009kk,Ammon:2009wc,Gursoy:2012bt,He:2013qq,Alho:2013hsa,Bigazzi:2014qsa,Jarvinen:2015ofa,Gursoy:2018ydr,Watanabe:2019zny} have been widely studied.
Aimed at better understanding the internal structure of hadrons, various holographic QCD models have been proposed and successfully implemented for high energy scattering~\cite{Polchinski:2001tt,Polchinski:2002jw,Brower:2006ea,Hatta:2007he,Pire:2008zf,Marquet:2010sf,Watanabe:2012uc,Watanabe:2013spa,Watanabe:2015mia,Watanabe:2018owy,Xie:2019soz,Liu:2022out,Zhang:2023nsk,Liu:2022zsa,Liu:2023tjr,Watanabe:2023rgp,Watanabe:2023kki}.
A holographic model inspired by the Regge theory and the structure of string scattering amplitudes was used to simulate scattering in the Regge regime, where $s \gg t$, for $pp$ and $p\Bar{p}$ and successfully described the experimental data for the scattering cross sections~\cite{Domokos:2009hm,Domokos:2010ma,Liu:2022zsa}.
The string dual models of QCD developed on the basis of the AdS/CFT correspondence allow the perturbative calculations of string theory to map the dynamics of nonperturbative strong interaction in the higher dimensional curved space.
Glueball states are dual to closed strings and meson states are dual to open strings, which correspond to the Pomeron and Reggeon exchange, respectively.
The leading Pomeron trajectory has an intercept slightly larger than 1, corresponding to the increasing behavior of the total cross section with the energy $\sqrt{s}$.
In contrast, the intercept of the Reggeon trajectory is smaller than 1, corresponding to the decreasing behavior of the total cross section.

For the Coulomb interaction, the electromagnetic effect -- soft photon radiation and Coulomb scattering -- is an integral element for any processes involving charged hadrons.
Occasionally this effect hampers the observation of specific strong interaction phenomena, but meanwhile they provide a unique source of information about important details of hadron amplitudes.
Within the leading approximation (one photon exchange), the Coulomb scattering amplitude is calculated in the framework of quantum electrodynamics (QED).
As described in Ref.~\cite{Kaspar:2011eva}, there are reasons to believe that this approximation is sufficient in the region of very low momentum transfer, where the Coulomb amplitude dominates.
Between the regions, in which the scattering is mostly Coulomb or mostly strong interaction, there is an interval of angles, where the two interactions have similar strengths.
The interference effect can be observed in this region, caused by the cross term of the Coulomb and strong interaction, the so-called Coulomb-nuclear interference (CNI).
This effect is the third contribution to the differential cross section in addition to the two fundamental interactions.
The CNI effect has been considered in experimental studies of $pp$ and $p\Bar{p}$ scattering to reveal the amplitude structure of hadron scattering~\cite{Petrov:2002nt,TOTEM:2013lle,ATLAS:2014vxr,TOTEM:2016lxj,Kohara:2017ats,TOTEM:2017sdy,TOTEM:2017asr,Khoze:2019fhx,Kaspar:2020oih,Grafstrom:2020zzg}.
This interference is clearly visible only in a very limited range of scattering angles.
It has been shown in Refs.~\cite{Amos:1985wx,UA4:1987gcp} that the contribution of the Coulomb interaction and its interference term with the strong interaction nearly vanishes at $|t| = 0.01$~GeV$^2$. 
However, analyzing the differential cross section in this interval can give us important information about the internal structure of hadrons.

In Ref.~\cite{Liu:2023tjr} a holographic QCD model, considering the Pomeron and Reggeon exchange contributions, has been employed for successfully describing the high energy $\pi p$ cross section data in the kinematic range of $0.01 < |t| < 0.45$~GeV$^2$ and $\sqrt{s} > 10$~GeV.
In their model, for the Pomeron-hadron couplings the gravitational form factors obtained from the bottom-up AdS/QCD models~\cite{Abidin:2008hn,Abidin:2009hr} were used.
The present work is an extension of Ref.~\cite{Liu:2023tjr} by additionally considering the Coulomb interaction as well as the interference effect with the strong interaction in $\pi p$ scattering. 
Since the observable range of these two contributions is very limited, we need to consider an even smaller range of momentum transfer, which we study in the range, $0 < |t| < 0.01$~GeV$^2$.
The Coulomb interaction is described in terms of the purely real photon exchange QED amplitude, and there have been various theoretical studies of the CNI effect~\cite{Bethe:1958zz,West:1968du,Franco:1973ei,Cahn:1982nr,Kundrat:2004ri,Kundrat:2007iv,Petrov:2018xma,Petrov:2022fsu}.
In addition, the electromagnetic form factors obtained from the bottom-up AdS/QCD models~\cite{Grigoryan:2007wn,Abidin:2009hr} are utilized.
By combining all the components, we derive an expression for the total scattering amplitude and numerically calculate the differential cross section.
We show that our predictions for the $\pi^+ p$ and $\pi^- p$ differential cross sections are consistent with the experimental data.
Besides, we explicitly exhibit how the contribution of each component varies with energy, focusing on the contribution ratio.

The composition of this paper is as follows.
In Sec.~\ref{sec:model} we derive the complete scattering amplitude and show each of the components.
In Sec.~\ref{sec:results} comparisons between our predictions for the differential cross sections and the experimental data are presented, and the $t$ and $\sqrt{s}$ dependence of each contribution are shown in detail.
A concluding remark is given in Sec.~\ref{sec:conclusion}.

\section{Model setup}
\label{sec:model}

\subsection{Holographic description of strong interaction amplitude}
\label{sec:Strong interaction}
In the preceding study~\cite{Liu:2023tjr}, the formalism for strong interaction which describes the $\pi p$ scattering with the Pomeron and Reggeon exchange was developed in the Regge regime.
According to the formalism, the strong interaction amplitude is obtained by combining the Pomeron and Reggeon exchange, which are described by the massive spin-2 glueball and the vector meson exchange, respectively.
The strong interaction amplitude for the $\pi p$ scattering can be written in the following form
\begin{equation}
    F_{N}^{\pi p}=F_{g}^{\pi p}+F_{v}^{\pi p},\label{Nuclear Amplitude}
\end{equation}
where $F_{g}^{\pi p}$ is the glueball exchange amplitude, and $F_{v}^{\pi p}$ is the vector meson exchange amplitude. The upper index `$\pi p$' stands for `$\pi^+ p$' or `$\pi^- p$'.

In the Regge limit, the proton-glueball-proton and pion-glueball-pion vertex can be expressed as
\begin{equation}
\Gamma_{g p p}^{\mu \nu}=\frac{i \lambda_{g p p} A_p(t)}{2}\left(\gamma^\mu p_p^\nu+\gamma^\nu p_p^\mu\right),
\end{equation}
\begin{equation}
\Gamma_{g \pi \pi}^{\mu \nu}=2 i \lambda_{g \pi \pi} A_\pi(t) p_\pi^\mu p_\pi^\nu,
\end{equation}
respectively, where $p_p$ and $p_\pi$ are the four-momenta of the corresponding hadrons, $\lambda_{g p p}$ and $\lambda_{g \pi \pi}$ are the coupling constants for the glueball exchange, and $A_p(t)$ and $A_\pi(t)$ represent the proton and pion gravitational form factor, respectively. 
Similarly, the proton-vector-proton and pion-vector-pion vertex can be written as
\begin{equation}
\Gamma_{v p p}^\mu=-i \lambda_{v p p} \gamma^\mu,
\end{equation}
\begin{equation}
\Gamma_{v \pi \pi}^\nu=-2 i \lambda_{v \pi \pi} p_\pi^\nu,
\end{equation}
respectively, where $\lambda_{v p p}$ and $\lambda_{v \pi \pi}$ are the coupling constants for the vector meson exchange.

The massive spin-2 glueball exchange amplitude is obtained by combining the proton-glueball-proton vertex, pion-glueball-pion vertex, and corresponding propagator. 
The vector meson exchange amplitude can be obtained in a similar way.
Propagators of the glueball and vector meson can be expressed as~\cite{Yamada:1982dx,Anderson:2016zon}
\begin{equation}
D_{\alpha \beta \gamma \delta}^g(k)=\frac{-i \left(\eta_{\alpha \gamma} \eta_{\beta \delta}+\eta_{\alpha \delta} \eta_{\beta \gamma}\right)}{2 (k^2+m_g^2)},
\end{equation}
\begin{equation}
D_{\mu \nu}^v(k)=\frac{i}{k^2+m_v^2} \eta_{\mu \nu},
\end{equation}
respectively, where $k^2=t$, $m_g$ and $m_v$ are the glueball and vector meson mass.
Hence amplitudes of the glueball and vector meson exchange can be written as
\begin{equation}
    F_{g}^{\pi p}=\Gamma_{g \pi \pi}^{\alpha \beta} \bar{u}_4 \Gamma_{g p p}^{\gamma \delta} u_2 D_{\alpha \beta \gamma \delta}^g (k),
\end{equation}
\begin{equation}
    F_{v}^{\pi p}=\Gamma_{v \pi \pi}^\mu \bar{u}_3 \Gamma_{v p p}^\nu u_1 D_{\mu \nu}^v(k),
\end{equation}
respectively.

Combining the above equations, the invariant strong interaction amplitude for the elastic $\pi p$ scattering is derived as
\begin{equation}
\begin{aligned}
      F_{N }^{\pi p}= \lambda_{g \pi \pi} \lambda_{g p p} A_\pi(t) A_p(t) s^2 \times \frac{1}{t-m_g^2}-2 \lambda_{v \pi \pi} \lambda_{v p p} s \times \frac{1}{t-m_v^2}.
\end{aligned}
\end{equation}
The differential cross section for strong interaction can be expressed as
\begin{equation}
\begin{aligned}
\frac{d \sigma_N}{d t} & =\frac{1}{16 \pi s^{2}}\left| F_N^{\pi p}(s, t)\right|^{2} \\
& =\frac{ \lambda_{g \pi \pi}^{2} \lambda_{g p p}^{2} A_\pi(t)^{2} A_p(t)^{2} s^2}{16 \pi\left|t-m_{g}^{2}\right|^{2}}-\frac{\lambda_{g \pi \pi} \lambda_{g p p} \lambda_{v \pi \pi} \lambda_{v p p} A_\pi(t) A_p(t) s}{4 \pi\left|t-m_{g}^{2}\right|\left|t-m_{v}^{2}\right|}+\frac{\lambda_{v \pi \pi}^{2} \lambda_{v p p}^{2}}{4 \pi\left|t-m_{v}^{2}\right|^{2}}.
\end{aligned}
\end{equation}
Here the invariant amplitude only contains the lightest states on the Pomeron and Reggeon trajectory. In order to include the higher spin states, the string excited states need to be considered. 
The higher states on the Pomeron and Reggeon trajectory correspond to the excited states of the closed and open string, respectively.
Following the preceding work~\cite{Anderson:2016zon}, the Reggeized Pomeron propagator is obtained as
\begin{equation}
    \frac{1}{t-m_g^2}\quad\to\quad\left(\frac{\alpha_g'}{2}\right)e^{-\frac{i\pi\alpha_g(t)}{2}}\frac{\Gamma\left[-{\chi_g}\right]\Gamma\left[1-\frac{\alpha_g(t)}{2}\right]}{\Gamma\left[-{\chi_g}-1+\frac{\alpha_g(t)}{2}\right]}\left(\frac{\alpha_g's}{2}\right)^{\alpha_g(t)-2},
\end{equation}
where $\chi_{g}={\alpha_g^{\prime}}{{m_p}^{2}+\frac{3}{2} \alpha_g(0)}-3 $, and $m_p$ is the proton mass. The propagator of Reggeon needs to be replaced by
\begin{equation}
    \dfrac{1}{t-m_v^2}\quad\to\quad\alpha_v'\:e^{-\frac{i\pi\alpha_v(t)}{2}}\:\sin\left[\dfrac{\pi\alpha_v(t)}{2}\right]\:(\alpha_v's)^{\alpha_v(t)-1}\:\Gamma[-\alpha_v(t)]\:.
\end{equation}

With the Reggeized Pomeron and Reggeon propagator introduced above, the invariant strong interaction amplitude for the $\pi p$ scattering is expressed as
\begin{equation}
\begin{aligned}
F_N(s, t)= & -s\lambda_{g \pi \pi} \lambda_{g p p} A_\pi(t) A_p(t)  e^{-\frac{i \pi \alpha_{g}(t)}{2}} \frac{\Gamma\left[-\chi_{g}\right] \Gamma\left[1-\frac{\alpha_{g}(t)}{2}\right]}{\Gamma\left[\frac{\alpha_{g}(t)}{2}-1-\chi_{g}\right]}\left(\frac{\alpha_{g}^{\prime} s}{2}\right)^{\alpha_{g}(t)-1} \\
& +2s \lambda_{v \pi \pi} \lambda_{v p p}  \alpha_{v}^{\prime} e^{-\frac{i \pi \alpha_{v}(t)}{2}} \sin \left[\frac{\pi \alpha_{v}(t)}{2}\right]\left(\alpha_{v}^{\prime} s\right)^{\alpha_{v}(t)-1} \Gamma\left[-\alpha_{v}(t)\right] .
\end{aligned}
\end{equation}
The above equation involves nine adjustable parameters in total, and all the values we used in this study are shown in Table~\ref{table}. 
\begin{table}[t!]
\centering
\caption{Parameter values for the strong interaction.}
\begin{tabular}{l c l} 
      \hline 
      Parameter & \ Value & \ \ \ \ \ Source\\
      \hline
      $\alpha_g(0)$ & \ 1.086 & \ \ \ \ \ Fit to $pp$($p\Bar{p}$) data at high energies~\cite{Xie:2019soz}\\
      $\alpha_{g}^{\prime}$ & \ 0.377 $\rm GeV^{-2}$ & \ \ \ \ \ Fit to $pp$($p\Bar{p}$) data at high energies~\cite{Xie:2019soz}\\
      $\lambda_{g p p}$ & \ 9.699 $\rm GeV^{-1}$ & \ \ \ \ \ Fit to $pp$($p\Bar{p}$) data at high energies~\cite{Xie:2019soz}\\
      $\alpha_v(0)$ & \ 0.444 & \ \ \ \ \ Fit to $pp$($p\Bar{p}$) data at medium energies~\cite{Liu:2022zsa}\\
      $\alpha_{v}^{\prime}$ & \ 0.925 $\rm GeV^{-2}$ & \ \ \ \ \ Fit to $pp$($p\Bar{p}$) data at medium energies~\cite{Liu:2022zsa}\\
      $\lambda_{v p p}$ & \ 7.742 & \ \ \ \ \ Fit to $pp$($p\Bar{p}$) data at medium energies~\cite{Liu:2022zsa}\\
      $\lambda_{g \pi \pi}$ & \ 3.361 $\rm GeV^{-1}$ & \ \ \ \ \ Fit to $\pi p$ data at medium energies~\cite{Liu:2023tjr}\\
      $\lambda_{v \pi^- \pi^-}$ & \ 4.528 & \ \ \ \ \ Fit to $\pi p$ data at medium energies~\cite{Liu:2023tjr}\\
      $\lambda_{v \pi^+ \pi^+}$ & \ 6.049 & \ \ \ \ \ Fit to $\pi p$ data at medium energies~\cite{Liu:2023tjr}\\
      \hline
\end{tabular}
\label{table}
\end{table}
Due to the universality of Pomeron and Reggeon, we can use the values determined in previous works. Three of those parameters, the intercept $\alpha_g(0)$, slope $\alpha_{g}^{\prime}$, and proton-glueball coupling constant $\lambda_{g p p}$, were obtained in the high energy $pp$($p\Bar{p}$) scattering considering only the Pomeron exchange~\cite{Xie:2019soz}. Another three parameters, the intercept $\alpha_v(0)$, slope $\alpha_{v}^{\prime}$, and proton-vector coupling constant $\lambda_{v p p}$, were determined in the medium energy $pp$($p\Bar{p}$) scattering considering both the Pomeron and Reggeon contribution~\cite{Liu:2022zsa}. For the other three parameters, pion-glueball coupling constant $\lambda_{g \pi \pi}$ and pion-vector coupling constants $\lambda_{v \pi^- \pi^-}$ and $\lambda_{v \pi^+ \pi^+}$, we utilize the results obtained in Ref.~\cite{Liu:2023tjr}.

\subsection{Proton and pion gravitational form factors}
\label{sec:gravitational}
To complete the description of the strong interaction amplitude presented in the previous subsection, it is necessary to specify the gravitational form factors of proton and pion, $A_p$ and $A_{\pi}$, which can be obtained from the bottom-up AdS/QCD models~\cite{Abidin:2009hr,Abidin:2008hn}. 
We employ the results derived with the hard-wall model, in which the AdS geometry is sharply cut off at the infrared (IR) boundary $z=z_0$. The metric of five-dimensional AdS space is given by
\begin{equation}
d s^2=g_{M N} d x^M d x^N=\frac{1}{z^2} \eta_{M N} d x^M d x^N, \quad \varepsilon<z<z_0\label{metric},
\end{equation}
where $\eta_{M N}=\operatorname{diag}(1,-1,-1,-1,-1)$. The fifth-coordinate $z$ runs from $\varepsilon \rightarrow 0$ which corresponds to the ultraviolet (UV) boundary.

For the proton gravitational form factor~\cite{Abidin:2009hr}, the proton is characterized by the solution of the five-dimensional Dirac equation. Coupling the Dirac field in the five-dimensional AdS space with a vector field, the classical action can be expressed as
\begin{equation}
\begin{aligned}
S_F=  \int d^{d+1} x \sqrt{g} \left(\frac{i}{2} \bar{\Psi} e_A^N \Gamma^A D_N \Psi-\frac{i}{2}\left(D_N \Psi\right)^{\dagger} \Gamma^0 e_A^N \Gamma^A \Psi-M \bar{\Psi} \Psi\right)\label{action},
\end{aligned}
\end{equation}
where $e_A^N=z \delta_A^N$ is the inverse vielbein, $D_N=\partial_N+\frac{1}{8} \omega_{N A B}\left[\Gamma^A, \Gamma^B\right]-i V_N$ represents the covariant derivative, and $M$ stands for the mass of the bulk spinor.
The field $\Psi$ is a solution of the Dirac equation
\begin{equation}
\left[i e_A^N \Gamma^A D_N - M\right] \Psi=0.
\end{equation}
The Dirac field can be expressed in terms of right-handed and left-handed spinors $\Psi_{R, L}=(1 / 2)\left(1 \pm \gamma^5\right) \Psi$. The normalizable ground state wave functions can be expressed after imposing boundary conditions as
\begin{equation}
\begin{aligned}
& \psi_L(z)=\frac{\sqrt{2} z^2 J_2\left(m_p z\right)}{z_0^p J_2\left(m_p z_0^p\right)}, \\
& \psi_R(z)=\frac{\sqrt{2} z^2 J_1\left(m_p z\right)}{z_0^p J_2\left(m_p z_0^p\right)}.
\end{aligned}
\end{equation}
The IR boundary parameter satisfies the condition $J_1 \left(m_p z_0^p\right)=0$, and is found to be $z_0^p =1 /(245~\mathrm{MeV})$.

The QCD energy-momentum tensor matrix element can be expressed in terms of three gravitational form factors~\cite{Pagels:1966zza}
\begin{equation}
\begin{aligned}
\left\langle p^{\prime}, s^{\prime}\left|T_{\mu \nu}(0)\right| p, s\right\rangle=\bar{u}\left(p^{\prime}, s^{\prime}\right) & {\left[A_p(Q) \frac{\gamma_\mu P_\nu+\gamma_\nu P_\mu}{2}\right.} \\
& +B_p(Q) \frac{i\left(P_\mu \sigma_{\nu \rho}+P_\nu \sigma_{\mu \rho}\right) q^\rho}{4 m_p} \\
& \left.+C_p(Q) \frac{\left(q_\mu q_\nu-\eta_{\mu \nu} q^2\right)}{m_p}\right] u(p, s),
\end{aligned}
\end{equation}
where $q=p^{\prime}-p$, $t=q^2$, $Q^2=-t$, and $P=\left(p+p^{\prime}\right) / 2$. 
Since in the Regge regime both the $B_p(Q)$ and $C_p(Q)$ contribution can be neglected, we only need to consider the contribution of $A_p(Q)$ involved term. 
The proton gravitational form factor is obtained as
\begin{equation}
A_p(Q^2)=\int_\epsilon^{z_0^p} d z \frac{1}{2 z^{3}} \mathcal{H}(Q, z)\left(\psi_L^2(z)+\psi_R^2(z)\right).
\end{equation}
In the hard-wall model, the bulk-to-boundary propagator as the solution to the linearized Einstein equation can be expressed as~\cite{Abidin:2008hn}
\begin{equation}
\mathcal{H}(Q, z)=\frac{1}{2} Q^2 z^2\left(\frac{K_1\left(Q z_0^p\right)}{I_1\left(Q z_0^p\right)} I_2(Q z)+K_2(Q z)\right).\label{propagator}
\end{equation}

For the pion gravitational form factor, the effective action for the meson fields on the five-dimensional AdS space is given by~\cite{Erlich:2005qh}
\begin{equation}
S_M=\int d^5 x \sqrt{g} \left[ \mathrm{Tr} \left\{|D X|^2+3|X|^2-\frac{1}{4 g_5^2}\left(F_L^2+F_R^2\right)\right\}\right],
\end{equation}
where $X(x, z)=\frac{1}{2} \mathbb{I}\left(m_q z+\sigma z^3\right) \exp \left(2 i t^a \pi^a\right)$ is the bulk field and its covariant derivative is given by $D^M X=\partial^M X-i A_L^M X+i X A_R^M$. In this study, for simplicity we only consider the chiral limit, i.e. the quark mass $m_q=0$ and the pion is massless.
Then the action containing the pion field $\pi$ and the axial-vector field $A$ up to the second order is obtained as
\begin{equation}
\begin{aligned}
S_A=  \int d^5 x \sqrt{g}\left[\frac{v(z)^2}{2} g^{M N}\left(\partial_M \pi^a-A_M^a\right)\left(\partial_N \pi^a-A_N^a\right) -\frac{1}{4 g_5^2} g^{K L} g^{M N} F_{K M}^a F_{L N}^a\right],
\end{aligned}
\end{equation}
where $F_{K M}^a=\partial_K A_M^a-\partial_M A_K^a$, and $A=\left(A_L-A_R\right) / 2$. 
Taking the variation of this equation on $A_M^a$, one can obtain the equation of motion. The pion wave function can be obtained as~\cite{Grigoryan:2007wn}
\begin{equation}
\begin{aligned}
\Psi(z)=  z \Gamma\left[\frac{2}{3}\right]\left(\frac{\beta}{2}\right)^{\frac{1}{3}} \times\left(I_{-\frac{1}{3}}\left(\beta z^3\right)-I_{\frac{1}{3}}\left(\beta z^3\right) \frac{I_{\frac{2}{3}}\left(\beta\left(z_0^\pi\right)^3\right)}{I_{-\frac{2}{3}}\left(\beta\left(z_0^\pi\right)^3\right)}\right),\label{wave}
\end{aligned}
\end{equation}
where $\beta=g_5 \sigma / 3$ with $g_5=\sqrt{2}\pi$.
The IR boundary parameter satisfies the condition $J_0 \left(m_\rho z_0^\pi \right)=0$, where $m_\rho$ is the $\rho$ meson mass, and is found to be $z_0^\pi =1 /(323~\mathrm{MeV})$~\cite{in26}.
The pion decay constant can be written as
\begin{equation}
f_\pi^2  =-\frac{1}{g_5^2}\left(\frac{1}{z} \partial_z \Psi(z)\right)_{z=\epsilon \rightarrow 0}.
\end{equation}
This formula, combined with Eq.~\eqref{wave}, relates the parameter $\sigma$ to the pion decay constant
\begin{equation}
f_\pi^2=3 \cdot 2^{1 / 3} \frac{\Gamma[2 / 3]}{\Gamma[1 / 3]} \frac{I_{2 / 3}\left(\beta\left(z_0^\pi\right)^3\right)}{I_{-2 / 3}\left(\beta\left(z_0^\pi\right)^3\right)} \frac{\beta^{2 / 3}}{g_5^2}.
\end{equation}
Comparing with the experimental data $f_\pi=131~\mathrm{MeV}$, we can extract the parameter $\sigma$, and it is found that $\beta^{1/3}=424~\mathrm{MeV}$.

In general, hadrons with spin-0 have two gravitational form factors~\cite{Abidin:2008hn}. Focusing on the relevant part of the action, the matrix element can be obtained in terms of two independent form factors
\begin{equation}
\left\langle\pi^a\left(p_2\right)\left|T^{\mu \nu}(0)\right| \pi^b\left(p_1\right)\right\rangle= \delta^{a b}\left[2 A_\pi(Q) p^\mu p^\nu+\frac{1}{2} C_\pi(Q)\left(q^2 \eta^{\mu \nu}-q^\mu q^\nu\right)\right],
\end{equation}
where $C_\pi(Q)$ can be represented with $A_\pi(Q)$ as $C_\pi\left(Q\right)=A_\pi\left(Q\right) / 3$. Therefore, it is sufficient to consider only $A_\pi(Q)$ for the pion gravitational form factor. Then we obtain the transverse traceless component of the stress tensor matrix element at the origin in coordinate space,
\begin{equation}
\left\langle\pi^a\left(p_2\right)\left|\hat{T}^{\mu \nu}(0)\right| \pi^b\left(p_1\right)\right\rangle= 2 \delta^{a b} A_\pi\left(Q^2\right)\left[p^\mu p^\nu+\frac{1}{12}\left(q^2 \eta^{\mu \nu}-q^\mu q^\nu\right)\right].
\end{equation}
Utilizing these expressions and combining with the bulk-to-boundary propagator, Eq.~\eqref{propagator}, the pion gravitational form factor is expressed as
\begin{equation}
A_\pi(Q^2)=\int_\epsilon^{z_0^\pi} d z \mathcal{H}(Q, z)\left(\frac{\left(\partial_z \Psi(z)\right)^2}{g_5^2 f_\pi^2 z}+\frac{v(z)^2 \Psi(z)^2}{f_\pi^2 z^3}\right)\label{pion gravitational}.
\end{equation}

\subsection{Proton and pion electromagnetic form factors}
\label{sec:electromagnetic}
Besides the gravitational form factors, to consider the Coulomb interaction, which is discussed in the next subsection, we also need to specify the electromagnetic form factors of the proton and pion.
In this study we employ those obtained in Refs.~\cite{Abidin:2009hr,Grigoryan:2007wn}.
Similar to the gravitational form factors, both the proton and pion electromagnetic form factors can be calculated using the hard-wall AdS/QCD model with the same metric, Eq.~\eqref{metric}.

For the proton electromagnetic form factor~\cite{Abidin:2009hr}, the electromagnetic current matrix element can be represented by two independent form factors
\begin{equation}
\left\langle p_2, s_2\left|J^\mu(0)\right| p_1, s_1\right\rangle=  u\left(p_2, s_2\right)\left(F_1(Q) \gamma^\mu
+F_2(Q) \frac{i \sigma^{\mu \nu} q_\nu}{2 m_n}\right) u\left(p_1, s_1\right),
\end{equation}
where $q=p_2-p_1$ and $Q^2=-q^2$. The proton current operator can be written in terms of isoscalar and isovector currents
\begin{equation}
J_{p, n}^\mu=\chi_i\left(\frac{1}{2} J_S^\mu \delta_{i j}+J_V^{a \mu} t_{i j}^a\right) \chi_j,
\end{equation}
where $\chi=(1,0)$ for the proton.

According to the AdS/CFT correspondence, the four-dimensional isoscalar $J_S^\mu$ and isovector $J_V^{a \mu}$ current operators correspond to the five-dimensional gauge field isoscalar and isovector part, respectively. 
Both the isoscalar and isovector current matrix elements can be extracted from the three-point function
\begin{equation}
\left\langle 0\left|\mathcal{T} \mathcal{O}_R^i(x) J^{a \mu}(y) \overline{\mathcal{O}}_R^j(w)\right| 0\right\rangle.
\end{equation}
The relevant term in the action, Eq.~\eqref{action}, that contributes to the three-point function is given by
\begin{equation}
S_F=\int d^5 x \sqrt{g} e^{-\Phi} \bar{\Psi} e_A^M \Gamma^A V_M \Psi.
\end{equation}
However, the above equation provides only the $F_1(Q)$ form factor. Hence one needs to add the following gauge invariant term to the action to obtain the both form factors
\begin{equation}
\eta_{S, V} \int d^5 x \sqrt{g} e^{-\Phi} i \frac{1}{2} \bar{\Psi} e_A^M e_B^N\left[\Gamma^A, \Gamma^B\right] F_{M N}^{(S, V)} \Psi,
\end{equation}
where $\eta_{S}$ and $\eta_{V}$ are determined by the experimental value of the proton magnetic moment and used for the isoscalar and isovector components, respectively.

The invariant functions are given by
\begin{equation}
    C_1(Q)=\int_\epsilon^{z_0^p} dz\dfrac{V(Q,z)}{2z^{3}}(\psi_L{}^2(z)+\psi_R{}^2(z)) ,
\end{equation}
\begin{equation}
    C_2(Q)=\int_\epsilon^{z_0^p} dz\frac{\partial_z V(Q,z)}{2z^{2}}(\psi_L{}^2(z)-\psi_R{}^2(z)) ,
\end{equation}
\begin{equation}
    C_3(Q)=\int_\epsilon^{z_0^p} dz\frac{2m_p V(Q,z)}{z^{2}}\:\psi_L(z)\psi_R(z),
\end{equation}
where $V(Q,z)$ is the vector bulk-to-boundary propagator for the hard-wall model and satisfies the equation
\begin{equation}
z \partial_z\left(\frac{1}{z} \partial_z V(Q, z)\right)=Q^2 V(Q, z)
\end{equation}
with boundary conditions $V(Q, 0)=1$ and $\partial_z V\left(Q, z_0\right)=0$. Its explicit form is expressed as
\begin{equation}
V(Q, z)=Q z\left(\frac{K_0\left(Q z_0\right)}{I_0\left(Q z_0\right)} I_1(Q z)+K_1(Q z)\right).
\end{equation}
The electric and magnetic form factor for the proton are given by
\begin{equation}
G_E(Q)=C_1(Q)+\eta_p C_2(Q)-\tau \eta_p C_3(Q),
\end{equation}
\begin{equation}
G_M(Q)=C_1(Q)+\eta_p C_2(Q)+\eta_p C_3(Q),
\end{equation}
respectively, where $\eta_p=\left(\eta_V + \eta_S\right) / 2=0.224$ and $\tau=Q^2/4 m_p^2$.
The effective electromagnetic form factor for the proton is obtained as
\begin{equation}
G_{p}(Q^2)=\sqrt{\frac{1}{1+\tau}\left[G_{\mathrm{E}}^2(Q)+\tau G_{\mathrm{M}}^2(Q)\right]}.
\end{equation}

\begin{figure}[t]
\centering
\includegraphics[width=0.55\textwidth]{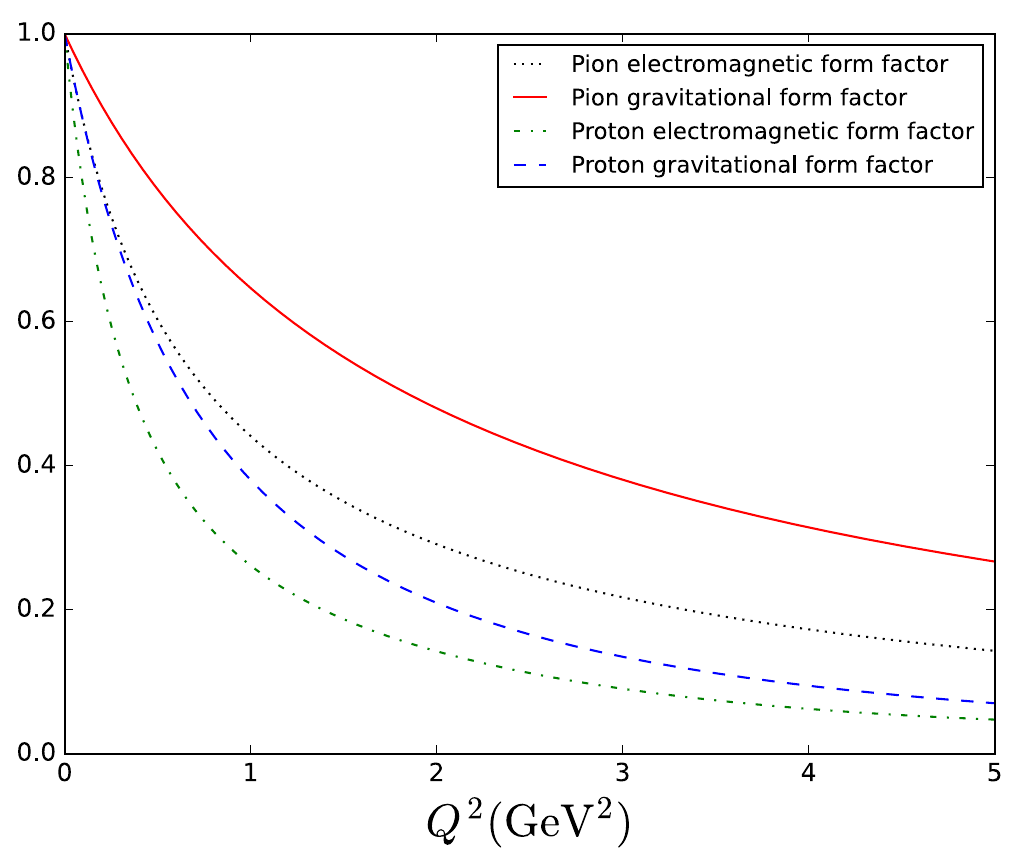}
\caption{Gravitational and electromagnetic form factors of the proton and pion as a function of $Q^2$. The solid and dotted curve represent results for the pion gravitational and electromagnetic form factor, respectively. The dashed and dash-dotted curve represent results for the proton gravitational and electromagnetic form factor, respectively.}
\label{four_form_factors}
\end{figure}

On the other hand, the pion electromagnetic form factor is derived as~\cite{Grigoryan:2007wn}
\begin{equation}
G_\pi(Q^2)=\int_\epsilon^{z_0^\pi} d z V(Q, z)\left(\frac{\left(\partial_z \Psi(z)\right)^2}{g_5^2 f_\pi^2 z}+\frac{v(z)^2 \Psi(z)^2}{f_\pi^2 z^3}\right).
\end{equation}
Note that, except for the bulk-to-boundary propagator $V(Q, z)$, this form factor is similar to the pion gravitational form factor introduced in the previous subsection. The $Q^2$ dependence of the four form factors is displayed in Fig.~\ref{four_form_factors}.
It can be seen that the proton form factors decrease faster than the pion form factors. Moreover, for the proton and pion the electromagnetic form factors decrease faster than the gravitational ones.

\subsection{Total scattering amplitude}
\label{sec:Coulomb interaction}
The elastic scattering of hadrons is realized mainly due to the strong interaction. 
However, in the scattering of charged hadrons, in addition to the strong interaction, the Coulomb interaction and CNI effect also exist simultaneously.
The influence of both the strong and Coulomb interaction in the $\pi p$ scattering is presently being described with the total elastic amplitude $F_{C+N}^{\pi p}$, which is written in the standard way as the sum of strong interaction amplitude $F_{N}^{\pi p}$ and Coulomb amplitude $F_{C}^{\pi p}$,
\begin{equation}
F_{C+N}^{\pi p}(s,t)=F_N^{\pi p}(s,t)+e^{i \alpha \phi(s,t)} F_C^{\pi p}(s,t),\label{tot}
\end{equation}
where $\alpha$ is the fine structure constant, and $\phi(s,t)$ is the Coulomb interference phase.

In this work the amplitude of Coulomb interaction is calculated in the leading approximation (one photon exchange) under the framework of QED, which can naturally explain the nucleon structure and its anomalous magnetic moments by introducing the electromagnetic form factor. 
The Coulomb interaction amplitude for pointlike charges considering the electromagnetic form factor can be expressed as
\begin{equation}
   F_C^{\pi p}(s,t)=\mp\dfrac{8\pi\alpha s}{|t|} G_{\pi} G_p ,
\end{equation}
in which the upper(lower) sign corresponds to the $\pi^+ p$($\pi^- p$) scattering.

Historically, there have been two ways to calculate the interference phase between the two fundamental interactions: the Feynman calculus or using the eikonal model. 
In the former one, the basic idea is to evaluate diagrams that include the Coulomb and strong interactions, and the result proposed by West and Yennie (WY)~\cite{West:1968du} is
\begin{equation}
\phi_{WY}(s, t)=\mp \left[\ln \left(\frac{-t}{s}\right)+\int_{-4 p^2}^0 \frac{\mathrm{d} t^{\prime}}{\left|t-t^{\prime}\right|}\left(1-\frac{F_{N}\left(s, t^{\prime}\right)}{F_{N}(s, t)}\right)\right].
\end{equation}
The complexity of this result is evident as it involves the integration over all permissible values of momentum transfer $t^\prime$. 
Furthermore, in WY's work the form of the strong interaction is not explicitly specified. 
For practical use, the authors employed a series of simplifying approximations to perform the analytical integration.  
The assumption has been made of a slow variation of the hadronic amplitude phase and a purely exponential decay of the hadronic amplitude. 
After adding some other high energy approximations and simplifications, the following simplified expression can be obtained
\begin{equation}
\phi_{WY}(s, t)=\mp \left[\ln \left(\frac{B(s) |t|}{2}\right)+\gamma\right],
\end{equation}
where $\gamma$ is the Euler constant, and $B(s)$ is the diffractive slope at $t=0$ generally expressed as
\begin{equation}
B(s)=\lim _{t \rightarrow 0} \frac{d\left[\ln \left(d \sigma_N / d t\right)\right]}{d t}.
\end{equation}

For the eikonal model~\cite{Franco:1973ei}, the scattering amplitude can be obtained via the Fourier-Bessel transform
\begin{equation}
F\left(s, q^2=-t\right)=\frac{s}{4 \pi i} \int d^2 b e^{i q b}\left[e^{2 i \delta(s, b)}-1\right],
\end{equation}
where $\delta(s, b)$ is the eikonal, and $b$ is the two-dimensional Euclidean vector.
Due to the additivity of corresponding potentials, the complete eikonal of two charged hadrons can be obtained by combining the Coulomb and hadronic eikonal
\begin{equation}
\delta_{\mathrm{tot}}(s, b)=\delta_C(s, b)+\delta_N(s, b).
\end{equation}
Using this assumption, the total scattering amplitude can be written as
\begin{equation}
F_{\mathrm{tot}}(s, q^2=-t)=  F_C(s, t)+F_N(s, t)  +\frac{i}{\pi s} \int d^2 q^{\prime} F_C\left(s, q^{\prime 2}\right) F_N\left(s,\left[q-q^{\prime}\right]^2\right)\label{eikonal}.
\end{equation}

The eikonal calculation has been investigated in various works~\cite{Franco:1973ei,Cahn:1982nr,Kundrat:2004ri,Kundrat:2007iv,Petrov:2018xma,Petrov:2022fsu}, and the commonly utilized eikonal model for the study of interference term begins with Eq.~\eqref{eikonal}. 
Based on this formula, R. Cahn made a series of simplifications and limited to small momentum transfer~\cite{Cahn:1982nr}. 
Cahn's results mainly tend to retrieve the formalism of WY on the basis of the eikonal model, taking into account the influence of the electromagnetic form factor, and a general expression for the Coulomb phase is obtained as
\begin{equation}
\phi_{Cahn}(s, t)=\mp \int_0^{\infty} \mathrm{d} t^{\prime} \ln \left(\frac{t^{\prime}}{t}\right) \frac{\mathrm{d}}{\mathrm{d} t^{\prime}}\left[f^2\left(t^{\prime}\right) \frac{F_N\left(s, t^{\prime}\right)}{F_N(s, 0)}\right].
\end{equation}
Similarly, by assuming that the hadronic amplitude undergoes the pure exponential decay with $t$ and the corresponding electromagnetic form factors are included in the derivation of the Coulomb phase, the following simplified form can be obtained
\begin{equation}
\begin{aligned}
\phi_{C a h n}= \mp[\ln \left(\frac{B|t|}{2}\right)+\gamma+C],\label{phase}
\end{aligned}
\end{equation}
\begin{equation}
    C=\ln \left(1+\frac{8}{B \Lambda^2}\right)+\left(4|t| / \Lambda^2\right) \ln \left(4|t| / \Lambda^2\right)+2|t| / \Lambda^2,
\end{equation}
where $\Lambda^2$ is a parameter involved in the dipole form factor $f(Q^2) = (1 + Q^2 / \Lambda^2)^{-2}$.
In the present study, we determine $\Lambda^2$ by comparing this dipole form factor with the electromagnetic form factor $G_p (Q^2) \times G_\pi (Q^2)$ which we previously introduced for the proton and pion.

In addition, within the leading approximation of $\alpha$, another significant work by Kundr\'{a}t and Lokaj\'{i}\v{c}ek (KL)~\cite{Kundrat:2004ri,Kundrat:2007iv} derived a closed-form interference formula. 
The complete form of the total scattering amplitude was derived using the eikonal model, instead of calculating the relative phase between the Coulomb and strong interaction. 
The main difference between KL's and Cahn's results lies in the derivation of Eq.~\eqref{eikonal}. 
Cahn directly simplified it, while KL took it as a whole and derived the complete scattering amplitude. 
However, in terms of numerical calculations, the both approaches demonstrated excellent consistency with experimental data, and there was virtually no difference.
According to Ref.~\cite{Kaspar:2020oih}, the results of WY exhibit an approximate 1\% deviation in numerical computations.
Furthermore, it is pointed out in Ref.~\cite{Kundrat:2007iv} that from the WY's result, high energy elastic hadron scattering is usually interpreted as central scattering and strongly limits the $t$ dependence of the hadron phase.
WY's integral formula has significant limitations on physical properties.

Recently, the TOTEM Collaboration~\cite{TOTEM:2017sdy} has used the eikonal model to extract the value of the $\rho$ parameter from $pp$ differential cross section data at the LHC energy. 
The simplified formula, Eq.~\eqref{phase}, was employed in our previous work~\cite{Zhang:2023nsk} to account for the Coulomb interaction in the $pp$ and $p\Bar{p}$ scattering within the framework of holographic QCD, and it was presented that the resulting differential cross sections are consistent with the data.
Hence, in this study we adopt the eikonal model.
Furthermore, since the strong interaction model presented in subsection~\ref{sec:Strong interaction} is highly complex, to simplify the numerical calculations we employ the simplified version, Eq.~\eqref{phase}, of Cahn's formula.

With the Coulomb interference phase introduced above, by combining the Coulomb and strong interaction amplitude and the corresponding form factors, the total amplitude can be obtained as
\begin{equation}
\begin{aligned}
 F_{C+N}(s, t)=& -e^{i \alpha \phi} \dfrac{8\pi\alpha s}{|t|} G_{\pi} G_{p} \\
 &-s\lambda_{g \pi \pi} \lambda_{g p p} A_\pi(t) A_p(t)  e^{-\frac{i \pi \alpha_{g}(t)}{2}} \frac{\Gamma\left[-\chi_{g}\right] \Gamma\left[1-\frac{\alpha_{g}(t)}{2}\right]}{\Gamma\left[\frac{\alpha_{g}(t)}{2}-1-\chi_{g}\right]}\left(\frac{\alpha_{g}^{\prime} s}{2}\right)^{\alpha_{g}(t)-1} \\
& +2s \lambda_{v \pi \pi} \lambda_{v p p}  \alpha_{v}^{\prime} e^{-\frac{i \pi \alpha_{v}(t)}{2}} \sin \left[\frac{\pi \alpha_{v}(t)}{2}\right]\left(\alpha_{v}^{\prime} s\right)^{\alpha_{v}(t)-1} \Gamma\left[-\alpha_{v}(t)\right],
\end{aligned}
\end{equation}
and the differential cross section is given by
\begin{equation}
\frac{d \sigma_{C+N}}{d t}=\frac{1}{16 \pi s^2}\left| F_{C+N}(s, t)\right|^2.
\end{equation}

\section{Numerical results}
\label{sec:results}
\subsection{Differential cross sections}
\label{sec:subsec_fitting_results}
In this section we numerically evaluate the parameterized form of the differential cross section for the $\pi^- p$ and $\pi^+ p$ scattering.
The strong interaction part involves nine relevant adjustable parameters, but the gravitational form factors do not bring any additional adjustable parameters.
For those nine parameters, as shown in Table~\ref{table}, we use the values obtained in the previous works~\cite{Xie:2019soz,Liu:2022zsa,Liu:2023tjr}.
In order to ensure the feasibility of the calculations, we choose the simplified form, Eq.~\eqref{phase}, for the interference phase.
Since the value of parameter $\Lambda^2$ in Eq.~\eqref{phase} depends on the involved hadron, it is necessary to newly determine $\Lambda^2$ which is related to the electromagnetic form factor.
Once this parameter is determined, the $\pi p$ differential cross sections can be calculated without any additional parameters.
In this study we determine this parameter using the pion and proton electromagnetic form factor in the range of $0 < |t| < 0.05$~GeV$^2$.
The best fit value for both the $\pi^- p$ and $\pi^+ p$ scattering is found to be: $\Lambda^2=1.02$~GeV$^2$.

\begin{figure}[t]
\centering
\includegraphics[width=0.79\textwidth]{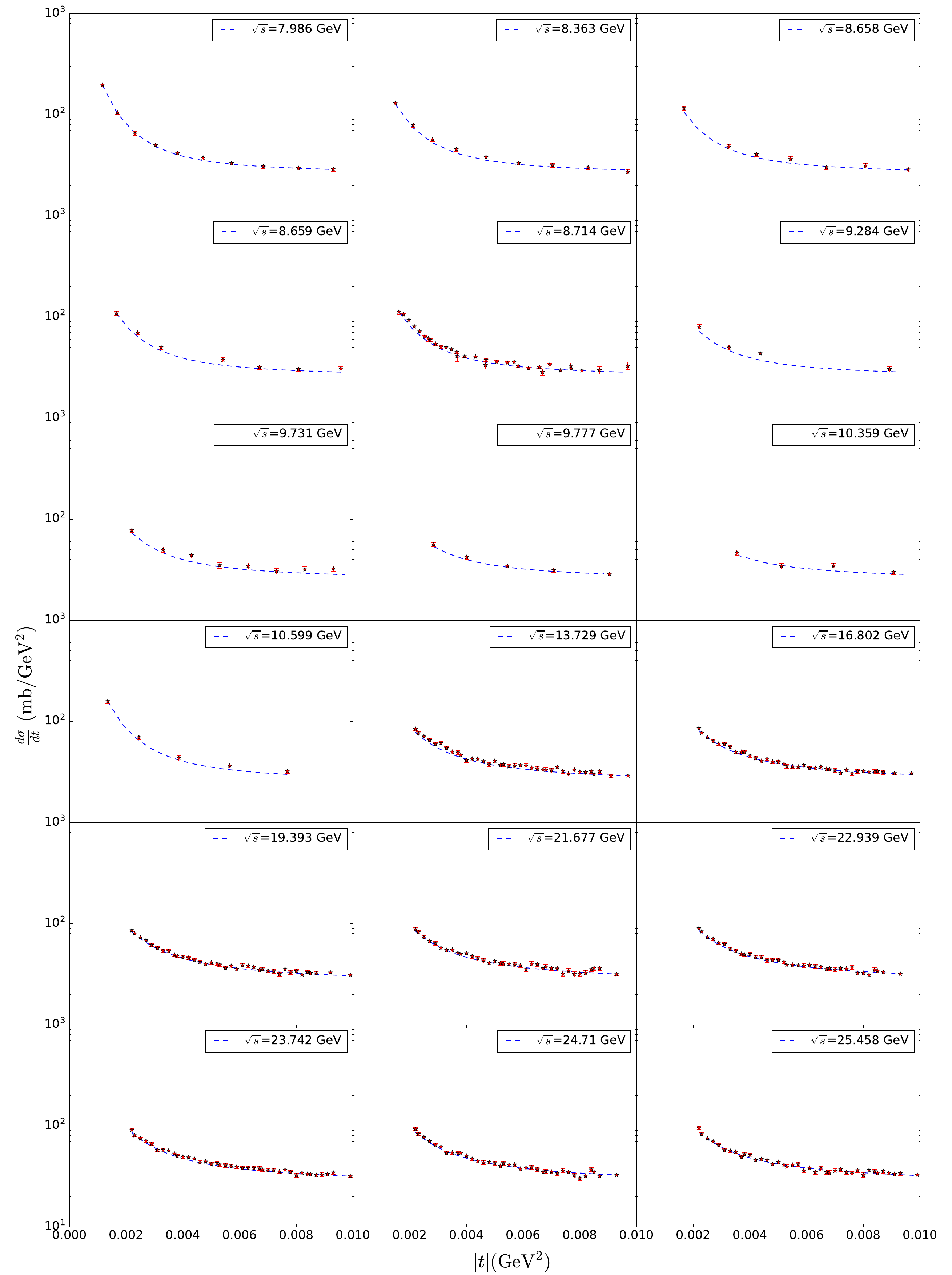}
\caption{
The differential cross section of the $\pi^- p$ scattering as a function of $|t|$.
The dashed curves represent our calculations.
The experimental data, represented by stars, are taken from Refs.~\cite{Apokin:1974mk,Apokin:1975ap,Apokin:1975rd,Apokin:1978sz,Burq:1982ja}.
}
\label{pi-p}
\end{figure}

\begin{figure}[t]
\centering
\includegraphics[width=0.6\textwidth]{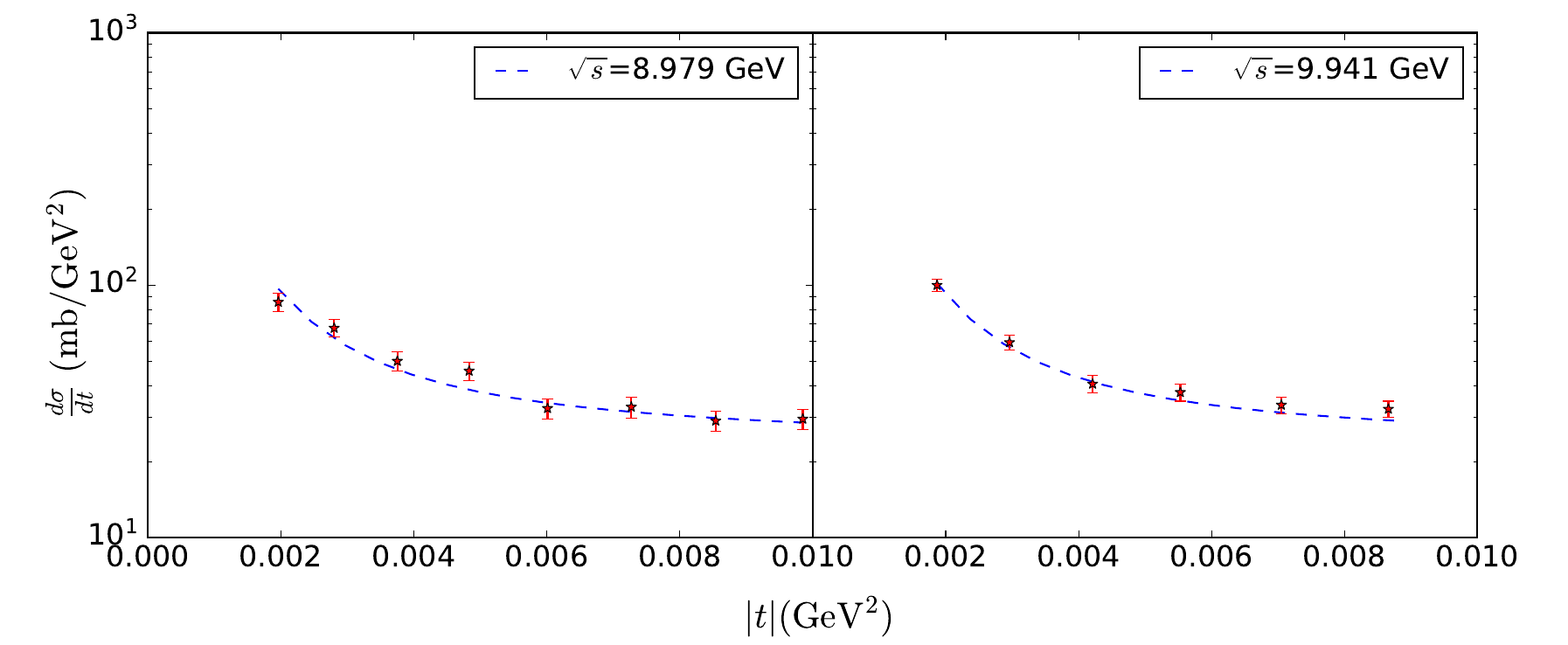}
\caption{
The differential cross section of the $\pi^+ p$ scattering as a function of $|t|$.
The dashed curves represent our calculations.
The experimental data, represented by stars, are taken from Ref.~\cite{Apokin:1976zu}.}
\label{pi+p}
\end{figure}

Since the chosen range of momentum transfer $|t|$ is quite small, in this study we consider the kinematic region with $\sqrt{s} \geq 5$~GeV, which satisfies the condition $s \gg t$.
We display in Fig.~\ref{pi-p}
our predictions for the $\pi^- p$ differential cross section, which are compared with the experimental data~\cite{Apokin:1974mk,Apokin:1975ap,Apokin:1975rd,Apokin:1978sz,Burq:1982ja}.
It can be seen that our predictions are overall consistent with the data in the considered kinematic region.
The results for the $\pi^+ p$ scattering are shown in Fig.~\ref{pi+p},
in which the experimental data are taken from Ref.~\cite{Apokin:1976zu}.
Although the available data are much less than those for the $\pi^- p$ case, it is found that our predictions agree with the data.

\subsection{Contribution ratios}
\label{sec:subsec_fitting_procedure}
\begin{figure}[t]
\centering
\begin{tabular}{ccccc}
\includegraphics[width=0.48\textwidth]{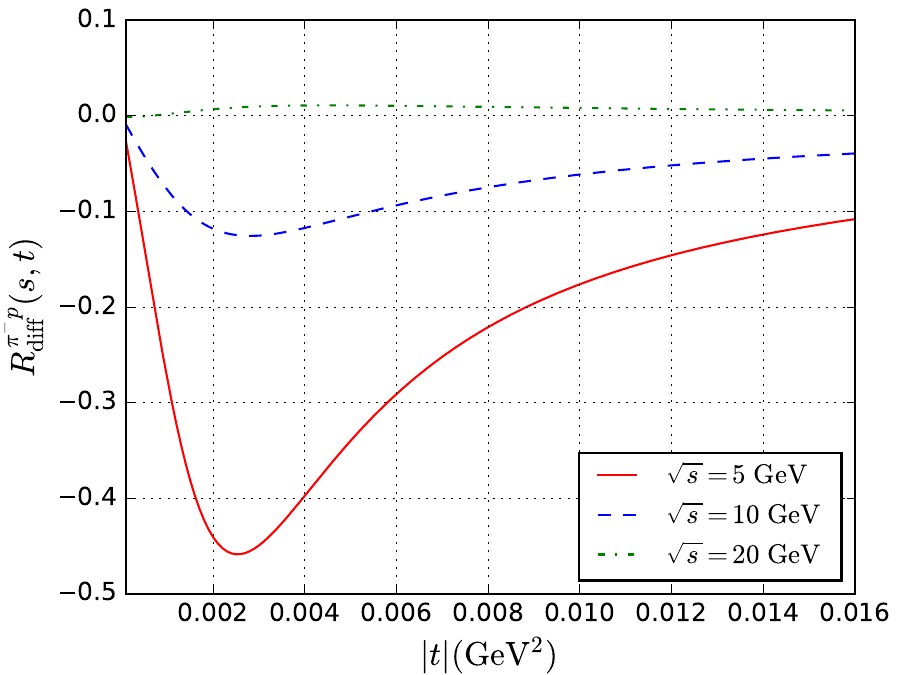}
\includegraphics[width=0.48\textwidth]{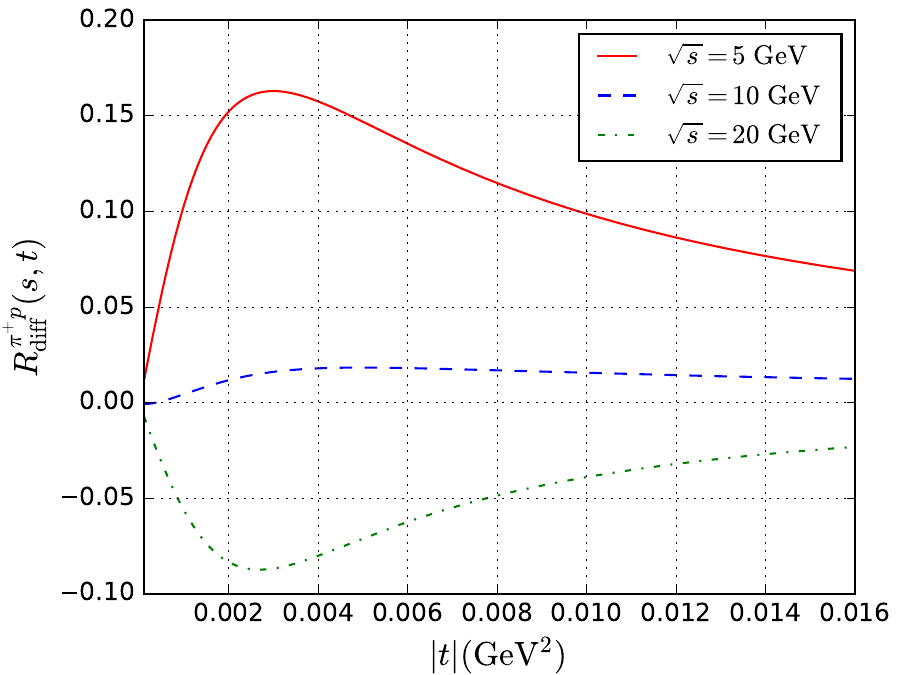}
\end{tabular}
\caption{
Contribution ratios of the interference effect in the $\pi^- p$ (left) and $\pi^+ p$ (right) differential cross section as a function of $|t|$ for $\sqrt{s}=5$, 10, 20~GeV.
}
\label{t_dependence}
\end{figure}

Here we investigate in detail the contribution of the interference effect between the Coulomb and strong interaction, defining the contribution ratio,
\begin{equation}
R=\frac{\frac{d \sigma_{C+N}}{d t}-\frac{d \sigma_{N }}{d t}-\frac{d \sigma_{C }}{d t}}{\frac{d \sigma_{C+N}}{d t}}.
\end{equation}
The resulting ratios for the $\pi^- p$ and $\pi^+ p$ scattering are shown in Fig.~\ref{t_dependence}.
From the results, strong $|t|$ and $\sqrt{s}$ dependence can be seen.
For the both $\pi^- p$ and $\pi^+ p$ scattering, the absolute value of the ratio becomes maximum around $t = 0.002$~GeV$^2$.
In the $\pi^- p$ case, the ratio increases with $\sqrt{s}$, and the ratios at $\sqrt{s}=5$, 10~GeV are negative in the considered $|t|$ range, while that at $\sqrt{s}=20$~GeV is positive but its magnitude is quite small.
In the $\pi^+ p$ case, the ratio decreases with $\sqrt{s}$, and the ratios at $\sqrt{s}=5$, 10~GeV are positive in the considered $|t|$ range, while that at $\sqrt{s}=20$~GeV is negative.
It is found that the $\pi^- p$ scattering at relatively low center-of-mass energy is more sensitive to the interference effect.

\begin{figure}[t]
\centering
\begin{tabular}{ccccc}
\includegraphics[width=0.48\textwidth]{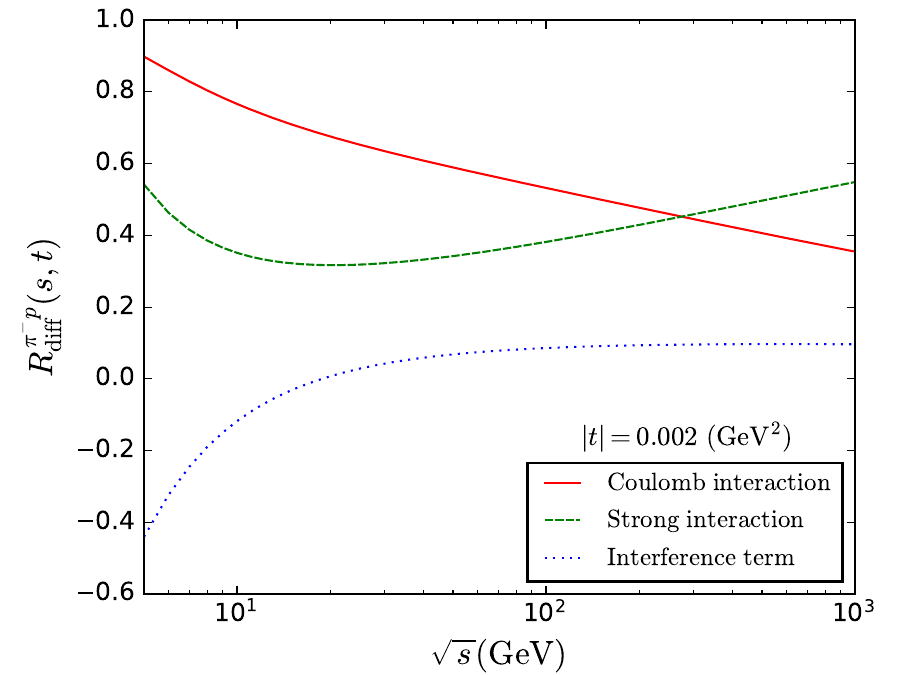}
\includegraphics[width=0.48\textwidth]{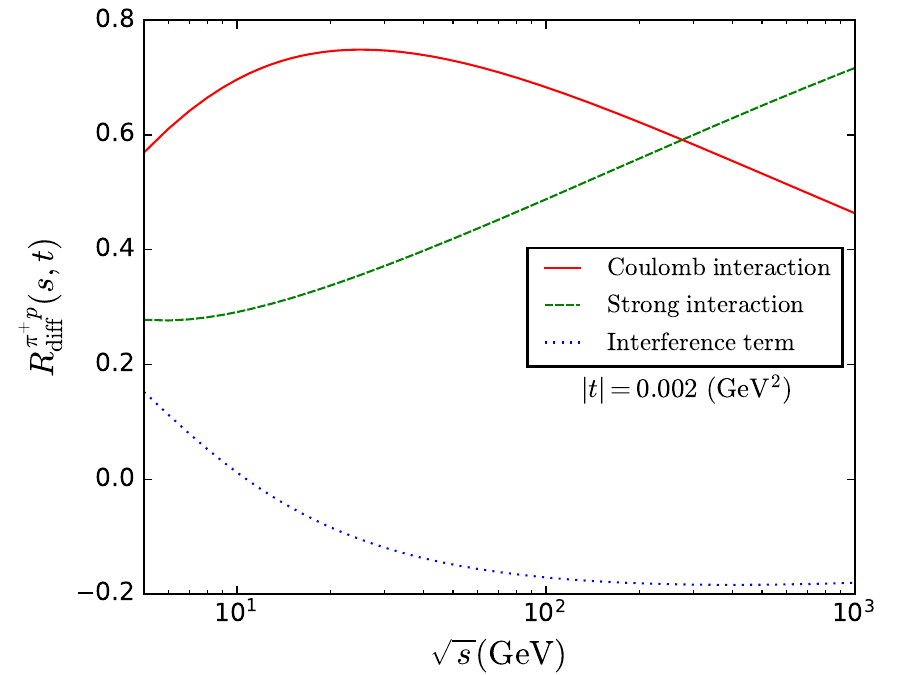}
\end{tabular}
\caption{
Contribution ratios in the $\pi^- p$ (left) and $\pi^+ p$ (right) differential cross section as a function of $\sqrt{s}$ at $|t|=0.002$~GeV$^2$.
The solid, dashed, and dotted curves represent results for the Coulomb interaction, strong interaction, and interference term, respectively.
}
\label{s_dependence}
\end{figure}

We also investigate ratios of the three contributions from the strong interaction, Coulomb interaction, and interference term in the $\pi^- p$ and $\pi^+ p$ differential cross section at $|t|=0.002$~GeV$^2$.
The resulting ratios are displayed in Fig.~\ref{s_dependence}.
From the results, it is seen that in this very low $|t|$ region the contribution from the Coulomb interaction, including the interference term, is substantial in the quite wide range of center-of-mass energy.
Since the currently available $\pi p$ differential cross section data concentrate in relatively narrow $\sqrt{s}$ range and there are no data at higher energies, it is expected that these nontrivial contribution ratios obtained in this study will be tested by future experiments.

\section{Conclusion}
\label{sec:conclusion}
We have studied the elastic $\pi p$ scattering in a holographic QCD model, focusing on the Regge regime.
To investigate differential cross sections of $\pi^- p$ and $\pi^+ p$ scattering in a wider kinematic region, we have taken into account both the strong and Coulomb interaction.
The strong interaction is realized by considering the Pomeron and Reggeon exchange, which are described by the Reggeized spin-2 glueball and vector meson propagator, respectively.
To obtain the strong interaction amplitude, we have combined the vertex factors, Reggeized propagators, and gravitational form factors derived with the bottom-up AdS/QCD model.
The Coulomb interaction amplitude is represented by the one photon exchange amplitude, utilizing the pion and proton electromagnetic form factor.
The total amplitude for the scattering process is not merely the addition of the two interactions.
There is the interference effect between the two interactions, which is expressed with the relative phase.
To obtain this phase, the eikonal model is adopted in this study.

There are nine adjustable parameters in the strong interaction part, but for all of them the values determined in the previous studies can be used.
There is one adjustable parameter in the interference term, and we have determined it using electromagnetic form factors of the pion and proton.
Once this parameter is determined, differential cross sections of the $\pi p$ scattering can be predicted without any additional parameters.
We have numerically evaluated the differential cross sections of the $\pi^- p$ and $\pi^+ p$ scattering, and shown that our predictions are consistent with the experimental data in all the considered kinematic region.
Furthermore, we have investigated in detail the $|t|$ and $\sqrt{s}$ dependence of the interference term and the ratios of each contribution in the differential cross sections.
To pin down those nontrivial quantities, more data especially at higher center-of-mass energies are necessary.
It is expected that future experiments will help to deepen our understanding of the Coulomb interaction in hadron-hadron scattering.

\begin{acknowledgments}
This work was supported in part by the National Natural Science Foundation of China under Grant Nos. 12175100 and 12375137, the Natural Science Foundation of Hunan Province of China under Grant No. 2022JJ40344, and the Research Foundation of Education Bureau of Hunan Province, China (Grant No. 21B0402).
\end{acknowledgments}

\bibliography{references}

\end{document}